\documentclass[preprint,authoryear,12pt]{elsarticle}
\usepackage{amsmath, amssymb}
\usepackage{dsfont}
\usepackage[latin1]{inputenc}
\usepackage[usenames]{color}
\usepackage[round]{natbib}
\usepackage{graphicx}
\usepackage{enumitem}
\usepackage{longtable} \usepackage{booktabs}
\usepackage{nomencl}

\newtheorem{theorem}{Theorem}
\newtheorem{lemma}{Lemma}
\newtheorem{corollary}[theorem]{Corollary}

\begin{document}

\begin{frontmatter}


\title{Maximum Likelihood Estimation of a Proportion from a Sample of Triplets}

\author{Rafael Wei\ss bach\corref{cor1}}
\ead{rafael.weissbach@uni-rostock.de}
\cortext[cor1]{Corresponding author: Prof. Dr. Rafael Wei\ss bach, Chair in Statistics and Econometrics,
Faculty for Economic and Social Sciences, University of Rostock, 18051 Rostock, Germany,
Phone: +49-381-4984428, Fax: +49-381-4984401.}
\author{Eric Scholz}
\address{Chair in Statistics and Econometrics, University of Rostock, Germany}

\begin{abstract}
When estimating a proportion and only a sample of triplets is given, dependencies within the triplets are to be accounted for. Without assuming a distribution for the success count of the triplet, together with the proportion, as second and third parameter the correlations of 1$^{st}$ and 2$^{nd}$ order enter the model. We apply maximum likelihood estimation, and derive consistency by using that the triplet count is multinomially distributed, combined with the continuous mapping theorem. The asymptotic normality follows with the delta-method, resulting in closed-form expressions for the standard errors. As application we study caries prevalence of pre-school children from a sample to nursing schools. We compare the standard errors with those for assuming erroneously independence within the nursing schools. As to be suspected, the design `inflates' the standard error markedly. 
\end{abstract}


\end{frontmatter}

\section{Introduction}
Occasionally, with interest on a proportion in a population, one may not draw units unrestrictedly as simple random sample (srs-design), but rather one has to draw clusters of units, of equal and unequal size (\citet[][Chapt. 9+9A]{cochran1977}, \citet[][Chapt. 4]{saerndal1992}, \citet[][]{donner2000}, \citet[see][]{bland2004}, \citet[][]{Wool2003}, or \citet[][Sect. IV.2.2]{lrh2020}). 

The intra-cluster correlation (ICC) can be of interest for its own, e.g. in finance when the cluster is a credit portfolio \cite[see e.g.][]{modelling-:2006}. More often the ICC is a nuisance parameter when estimating the proportion (or regression models build thereon \cite[see e.g.][Chapt. 8]{saerndal1992}).
The design effect is long-known that of 'inflating' the variance (VIF), and hence the standard error and the confidence interval, of the proportion \cite[see e.g.][]{Lyn}. When cluster are larger than pairs \cite[for siblings see e.g.][]{weissbach2013}, correlation between more than two units are to be accounted for \cite[see e.g.][]{stefanescu2003}. Here we study the case of three, i.e. for triplets, for a dichotomous outcome and apply maximum likelihood theory. In detail we use the Bahadur representation to parametrize the count of positive outcomes per triplet as a multinomial random variable, i.e. as a member of the Exponential family. We derive consistency thereof in a simplified proof and by the delta-method we derive asymptotic normality and the asymptotic variance of the estimator for the population proportion, and the correlations of 1$^{st}$ and 2$^{nd}$ order. 

The sequence of likelihoods, seen as functions in the parameter $\pmb{\rho}$, converges point-wise to the Kulback-Leibler loss by the LLN. When convergence is also uniform, the maxima of the functions will converge against the minimum of the loss, being the true parameter. Hence, for the consistency proof, we devote effort to establish uniform bounds.

Results in Section \ref{supersec1} are the closed-form expressions for the three maximum likelhood estimators in Subsection \ref{sec1}. In Subsection \ref{sec2} their asymptotic normality is shown following closely \citet[][Chapt. 5]{vaart1998}. In Section \ref{sec3}, after a short Monte Carlo simulation, we apply the confidence interval to the question of caries prevalence in the state of Hesse in Germany from a sample of nursing schools and compare it with the srs-design, to find that the VIF increases the confidence interval width by 10\%.  

\section{Result of maximum likelihood estimation} \label{supersec1}
 
\subsection{Notation, parametrizations, estimators} \label{sec1}
Let $(X_{i1}, X_{i2}, X_{i3})^T, \, i \in \{1, \ldots, n\}$, be independent triplets of exchangeable binary random variables, with $X_{ik}=1$ (success) or $X_{ik}=0$ (failure), $k \in \{1,2,3\}$. We assume that

\begin{enumerate}[label= \textnormal{(A\arabic{*})}, leftmargin=1cm]
	\item \label{A1:Assumption1} $X_{ik}$ are identically distributed with $P\{X_{ik}=1\}=\pi \in [ \xi_{\pi}, 1-\xi_{\pi}]$, where $\xi_{\pi}>0$ is arbitrary small, for $i \in \{1, \ldots, n\}$, $k \in \{1,2,3\}$.
\end{enumerate} 

Note that the random variables within a triplet are not assumed to be independent. We define the model in Bahadur representation consisting of three probabilities $\pmb{q}:=(q_1, q_2, q_3)^T \in [\xi, 1-\xi]^3$ with some small $\xi<1/3$ and $q_s:= P[S_i=s]$ where $S_i:=\sum_{k=1}^3 X_{ik}, \, s \in \{0,1,2,3\}$.

 Note that the numbers of triplets with $s$ success are multinomial distributed, i.e. with probability mass distribution (see e.g. \citet[][Chap. 35]{johnson} or \citet[][Sect. 9.6]{Bley2021}):
\begin{multline*}
 f_M\left(\sum_{i=1}^n\mathbb{I}_{\{s_i=0\}},\sum_{i=1}^n\mathbb{I}_{\{s_i=1\}},\sum_{i=1}^n\mathbb{I}_{\{s_i=2\}},\sum_{i=1}^n\mathbb{I}_{\{s_i=3\}}/ n, q_0, q_1,q_2,q_3\right) := \\
 n! \left[\left(\sum_{i=1}^n\mathbb{I}_{\{s_i=0\}}\right)! \ldots \left(\sum_{i=1}^n\mathbb{I}_{\{s_i=3\}}\right)!\right]^{-1} q_0^{\sum_{i=1}^n\mathbb{I}_{\{s_i=0\}}} \ldots q_3^{\sum_{i=1}^n\mathbb{I}_{\{s_i=3\}}}
\end{multline*} 
 with $\sum_{s=0}^3 \sum_{i=1}^n\mathbb{I}_{\{s_i=s\}} = n$ and  $\sum_{s=0}^3 q_s = 1$.  Hence, the maximum likelihood estimator $\hat{\pmb{q}}_{n}$ has coordinates $\hat{q}_{n,h}:= n^{-1}\sum_{i=1}^n \mathbb{I}_{\{S_i=h\}}$, for $h \in \{1,2,3\}$.

By the invariance property of maximum likelihood estimation, it carries over to the second parametrization, called $\pmb{\rho}$-parametrization, consisting of $\pi$, the well-known pairwise correlation $\rho_1$ and the correlation of second order $\rho_2:=\mathrm{E}[(X_{1} - \mathrm{E}(X_{1}))(X_2 - \mathrm{E}(X_2))(X_3- \mathrm{E}(X_3))]/[\pi(1-\pi)]^{3/2}$,  
where we suppress the subscript $i$. By interchangeability we can simplify to
\begin{align*}
	\rho_1=\frac{\mathrm{E}(X_1 X_2) -\pi^2 }{\pi(1-\pi)} \quad \text{and} \quad 
	\rho_2=\frac{\mathrm{E}(X_1 X_2 X_3)-3 \pi \mathrm{E}(X_1 X_2) + 2 \pi^3 }{[\pi(1-\pi)]^{3/2}},
\end{align*}
where $\mathrm{E}(X_1 X_2)=P\{X_1=1, X_2=1\}$. Note that, in terms of the Bahadur representation  \cite[see e.g.][]{stefanescu2003}, we have 
$\pi = 3^{-1} \sum_{s=0}^3 s q_s $, 
$\mathrm{E}(X_1 X_2)= 6^{-1} \sum_{s=2}^3 s(s-1) q_s $ and 
$\mathrm{E}(X_1 X_2 X_3) = q_3$, 
so that we can interpret $\pmb{\rho}$ as a function $\pmb{\phi}(\pmb{q})$ with

\begin{align}
	\pmb{\rho}:=\begin{pmatrix} \pi \\ \rho_1 \\ \rho_2 \end{pmatrix} = \begin{pmatrix} \phi_1(\pmb{q}) \\ \phi_2(\pmb{q}) \\ \phi_3(\pmb{q}) \end{pmatrix}
	 = \pmb{\phi}(\pmb{q}):=\begin{pmatrix}
		\frac{1}{3}q_1+\frac{2}{3}q_2+q_3 \\
		\\
		\frac{q_2/3+q_3-(q_1/3+ 2 q_2/3+q_3 )^2}{(q_1/3+2 q_2/3+q_3) [1-(q_1/3+ 2 q_2/4+q_3 )]} \\
		\\
		\frac{q_3-(q_1+2q_2+3q_3)(q_2/3+q_3)+2(q_1/3+2 q_2/3+q_3)^3}{\{[(q_1/3+ 2 q_2/3+q_3 )(1-(q_1/3+ 2 q_2/3+q_3 )]\}^{3/2}} 
	\end{pmatrix}. \label{Bahadur}
\end{align}
Note that, in contrast to the enumeration of the coordinates in $\pmb{q}$, the second coordinate of $\pmb{\rho}$ is $\rho_1$.

The idea is now to use consistency and asymptotic normality for the Bahadur representation and afterwards defer from them asymptotic results for $\hat{\pmb{\rho}}_n:= \pmb{\phi}(\hat{\pmb{q}}_n)$. From now on we write short $X_i^s := \mathbb{I}_{\{S_i=s\}}$ for $s \in \{1,2,3\}$ and $\pmb{X}_i:=(X_i^1, X_i^2, X_i^3)^T$. (Note that $\pmb{X}_i \neq (X_{i1}, X_{i2}, X_{i3})^T$.) Then, divided by $n$, the logarithmic likelihood for $\pmb{q}$ is 
\[
M_n(\pmb{q}):=\frac{1}{n} \log f_M \left( n-\Bigl(\sum_{i=1}^nX_i^1+X_i^2+X_i^3 \Bigr), \sum_{i=1}^n \pmb{X}_i / n, q_0,\pmb{q} \right).
\]
Thus, the estimator $\hat{\pmb{q}}_n$ will, (nearly) satisfy the system of equations $\pmb{\Psi}_n(\pmb{q}):=\frac{1}{n} \sum_{i=1}^n \pmb{\psi}_{\pmb{q}}(\pmb{X}_i)=\mathbf{0}$, 
with $\pmb{\Psi}_n:= \partial M_n(\pmb{q}) / \partial \pmb{q}$ \cite[notation as in][Chap. 5]{vaart1998}. The coordinates of $ \pmb{\psi}_{\pmb{q}}$ are, for $h \in \{1,2,3\}$, 
\begin{align}
	\psi_{\pmb{q}, h}(\pmb{X}_i):= \frac{1}{q_h(1-\sum_{j=1}^3q_j)}\left[\left(1-\sum_{\substack{j=1 \\ j \neq h}}^3 q_j\right)X_i^h+q_h\left(\sum_{\substack{j=1 \\ j \neq h}}^3 X_i^j -1\right) \right]. \label{psiqs}
\end{align}

\subsection{Large sample properties and standard error} \label{sec2}

A useful prerequisite for asymptotic normality is consistency. The multinomial distribution is a member of the Exponential family, so that $\hat{\pmb{q}}_{n}$ is consistent \cite[see e.g.][Example 1.6.7+Theorem 5.2.2(i)]{bickel}. Here we display a short and elementary proof and limit ourselves to the case of non-void categories, $\sum_{i=1}^n X_i^s>0$, for all $s \in \{0,\ldots,3\}$.
Note that the true value $\pmb{q}_0=(q_{0,1}, q_{0,2}, q_{0,3})^T$ is a zero of $\pmb{\Psi}(\pmb{q}):=\mathrm{E}_{\pmb{q}_0} \pmb{\psi}_{\pmb{q}} (\pmb{X}_i)$ with coordinates
\begin{align}
	\mathrm{E}_{\pmb{q}_0} \psi_{\pmb{q}, h}(\pmb{X}_i) =\frac{q_{0,h}}{q_h} - \frac{1- \sum_{j=1}^3q_{0,j}}{1- \sum_{j=1}^3q_{j}}, \quad h \in \{1,2,3\}. \label{expect}
\end{align}	

\begin{theorem} \label{theo1}
	Assuming independence between triplets and \ref{A1:Assumption1}, $\hat{\pmb{q}}_{n} \stackrel{n \to \infty}{\longrightarrow} \pmb{q}_0$ in probability.  
\end{theorem}

{\bf Proof.} We apply Lemma 5.10 of \cite{vaart1998} in its multivariate form.
	First of all, we have to show that $\pmb{\Psi}_n(\pmb{q}) \to \pmb{\Psi}(\pmb{q})$ in probability for every $\pmb{q}$.  Therefore, since the random vectors $\pmb{X}_i$ are independent and identically distributed, $\pmb{\psi}_{\pmb{q}}(\pmb{x})$ is continuous for almost all $\pmb{x}$, and the parameter space $\pmb{Q} \subset [\xi, 1-\xi]^3$ is compact, we can apply the uniform law of large numbers \cite[see][]{newey1994}. We only need to show that $\lVert \pmb{\psi}_{\pmb{q}}(\pmb{x})\rVert \leq g(\pmb{x})$ for almost all $\pmb{q} \in \pmb{Q}$ and a mapping $g(\pmb{x})$ with $\mathrm{E}_{\pmb{q}_0} g(\pmb{X}_i)<\infty$. This is fulfilled - with proof in \ref{appa} - if we choose
	\begin{equation} \label{hdef}
	g(\pmb{x}):= \sqrt{3} (1- \xi)(x_1+x_2+x_3)\xi^{-2}.
	\end{equation}
		Next, we have to verify that each map $\pmb{q} \mapsto \pmb{\Psi}(\pmb{q})$ is continuous and has exactly one zero $\hat{\pmb{q}}_n$. Because of non-void categories, this condition is fulfilled. 
	
	The proof is complete if $\pmb{q}_0$ is a point with $\pmb{\Psi}(\pmb{q}_0-\pmb{\epsilon}) >0 > \Psi(\pmb{q}_0+\pmb{\epsilon})$ for every $\pmb{\epsilon}=(\epsilon, \epsilon, \epsilon)^T >\pmb{0}$. By using \eqref{expect}, this can be verified directly. For example, it applies:
	\[
	\pmb{\Psi}_h(\pmb{q}_0-\pmb{\epsilon})=\mathrm{E}_{\pmb{q}_0} \psi_{\pmb{q}_0-\pmb{\epsilon}, h}(\pmb{X}_i)= \frac{q_{0,h}}{q_{0,h}-3\epsilon}-\frac{1-q_{0,1}-q_{0,2}-q_{0,3}}{1-q_{0,1}-q_{0,2}-q_{0,3}+3\epsilon}>0
	\] \qed
	
We may now use the continuous mapping theorem and the Bahadur representation \eqref{Bahadur}.
\begin{lemma} \label{lamm1}
	Under \ref{A1:Assumption1} and denoting by $\pmb{\rho}_0=\pmb{\phi}(\pmb{q}_0)$ the true parameter, with \eqref{Bahadur}, the sequence $\hat{\pmb{\rho}}_n=\pmb{\phi}(\hat{\pmb{q}}_{n})\stackrel{n \to \infty}{\longrightarrow} \pmb{\rho}_0$ in probability. 
\end{lemma}
Note that the point estimate of $\pi_0$ is simply the proportion,
\begin{equation} \label{pigleichpi}
	\hat{\pi}_n = \frac{1}{3} \sum_{s=0}^3 s \hat{q}_{n,s}= \frac{1}{3} \sum_{i=1}^n \sum_{h=1}^3 h \mathbb{I}_{\{S_i=h\}}  = \frac{1}{3n} \sum_{i=1}^n \sum_{k=1}^3 X_{ik},
\end{equation}
which is the same as for the na\"ive assumption of simple random sampling (srs-design), i.e. of intra-triplet independence between $X_{ik}$ and $X_{ik'}$. For the now $3 n$ random draws, one would have  $X_{11}, X_{12}, X_{13}, X_{21}, \ldots, X_{n 3} \stackrel{iid} \sim B(\pi_0)$ with maximum likelihood estimate $\hat{\pi}_n$. However, we will see later that the standard error is different.
\begin{corollary} \label{Cor1} With $\pmb{\phi}(\pmb{q})$ from \eqref{Bahadur} and $\hat{\pmb{\rho}}_n=\pmb{\phi}(\hat{\pmb{q}}_{n})$, it is 
	$\sqrt{n}(\hat{\pmb{\rho}}_n- \pmb{\rho}_0)$ asymptotically normal with mean zero and covariance matrix $\tilde{\pmb{\Sigma}}= \pmb{\phi}_{\pmb{q}_0}' \pmb{\Sigma} \pmb{\phi}_{\pmb{q}_0}'^T$, where
	\begin{align*}
		\pmb{\Sigma}:=\begin{pmatrix}
			q_{0,1}(1-q_{0,1}) & -q_{0,1}q_{0,2} & -q_{0,1}q_{0,3} \\
			-q_{0,1}q_{0,2} & q_{0,2}(1-q_{0,2}) & -q_{0,2}q_{0,3} \\
			-q_{0,1}q_{0,3} & -q_{0,2}q_{0,3} & q_{0,3}(1-q_{0,3})
		\end{pmatrix}
	\end{align*}
	and the components of $\pmb{\phi}_{\pmb{q}}':=(\frac{\partial \phi_{l_1}}{\partial q_{l_2}}(\pmb{q}))_{\stackrel{l_1 \in \{1,2,3\}}{l_2 \in \{1,2,3\}}}$ are given in closed-form as $\partial \phi_1(\pmb{q}) /\partial q_1=1/3, \partial \phi_1(\pmb{q})/\partial q_2=2/3, \partial \phi_1(\pmb{q}) \partial q_3=1, \partial \phi_2(\pmb{q})/\partial q_1 = - 3(q_{1}^2-9q_{3}^2+3q_{0,2}+9q_{3}+2q_{1}q_{2}-6q_{2}q_{3})(q_{1}+2q_{2}+3q_{3})^{-2}(3-q_{1}-2q_{2}-3q_{3})^{-2},\partial \phi_2(\pmb{q})/\partial q_2 =-3(3q_{1}^2+4q_{2}^2-9q_{3}^2-3q_{1}+9q_{3}+8q_{1}q_{2}+6q_{1}q_{3})(q_{1}+2q_{2}+3q_{3})^{-2}(3-q_{1}-2q_{2}-3q_{3})^{-2},  \partial \phi_2(\pmb{q})/\partial q_3 = -9(2q_{1}^2+4q_{2}^2-3q_{1}-3q_{2}+6q_{1}q_{2}+6q_{1}q_{3}+6q_{2}q_{3})(q_{1}+2q_{2}+3q_{3})^{-2}(3-q_{1}-2q_{2}-3q_{3})^{-2}, \partial \phi_3(\pmb{q})/\partial q_1=54^{-1}[(q_{1}/3+2 q_{2}/3+q_{3} )(1-q_{1}/3-2 q_{2}/3-q_{3} )]^{-5/2} [2q_{1}^3+(6q_{3}+8q_{2})q_{1}^2+(8q_{2}^2+3q_{2}-18q_{3}^2+27q_{3})q_{1} -54q_{3}^3+(81-72q_{2})q_{3}^2+(-24q_{2}^2+63q_{2}-27)q_{3}+6q_{2}^2], \partial \phi_3(\pmb{q})/\partial q_2= 27^{-1}[(3 q_{1}/3+2 q_{2}/3+q_{3} )(1-q_{1}/3- 2 q_{2}/3-q_{3} )]^{-5/2}
		[ 8 q_{2}^3+(12 q_{3}+20q_{1}-6)q_{2}^2+3q_{1}^3-3q_{1}^2 -27q_{3}^3 +(14q_{1}^2-9q_{1}-18q_{3}^2+27q_{3}+36q_{1}q_{3})q_{2}  +(9q_{1}+54)q_{3}^2+(15q_{1}^2+9q_{1}-27)q_{3}]$ and $
		\partial \phi_3(\pmb{q})/\partial q_3=18^{-1}[(q_{1}/3+ 2 q_{2}/3+q_{3} )(1- q_{1}/3- 2 q_{2}/3-q_{3} )]^{-5/2} [ (36q_{1}+36q_{2}+9)q_{3}^2+16q_{2}^3+(32q_{1}-26)q_{2}^2 +(20q_{1}^2-29q_{1}+12)q_{2}+4q_{1}^3-8q_{1}^2+6q_{1}  +(48q_{2}^2+72q_{1}q_{2}-33q_{2}+24q_{1}^2-21q_{1}-9)q_{3}]$.
		
\end{corollary}

{\bf Proof.} We apply Theorem 5.41 of \cite{vaart1998}. Note, that $\pmb{Q} \subseteq [\xi, 1-\xi]^3$ is a subset of the $3$-dimensional Euclidean space. 
	The derivative $\dot{\pmb{\psi}}_{\pmb{q}} (\pmb{X}_i)$ is a $3 \times 3$-matrix with the entries $\psi_{\pmb{q},h}^k = \partial \psi_{\pmb{q},h}/\partial q_k$. While the elements on the main diagonal satisfy
		\begin{multline}\label{deriHD}
		\psi_{\pmb{q},h}^h(\pmb{X}_i)=q_h^{-2}(1-\sum_{j=1}^3 q_j)^{-2} [ (1-\sum_{\substack{j=1 \\ j \neq h}}^3 q_j)X_i^h (q_s-1+\sum_{j=1}^3q_j)+q_h^2(\sum_{\substack{j=1 \\ j \neq h}}^3 X_i^j-1)], 
	\end{multline}
	the off-diagonal elements are determined by 
		\begin{align}
		\psi_{\pmb{q},h}^k(\pmb{X}_i)= \frac{X_i^1+X_i^2+X_i^3-1}{(1-q_1-q_2-q_3)^2}, \quad \text{for} \quad k \neq h. \label{deriND}
	\end{align}
The second order derivate $\ddot{\pmb{\psi}}_{\pmb{q}}(\pmb{X}_i)$ is described by a $ 3 \times 3 \times 3$-tensor, whose diagonal entries are given by:
		\begin{align*}
		\psi_{\pmb{q},h}^{h,h}(\pmb{X}_i) = \frac{1}{q_h^3(1-\sum_{j=1}^3 q_j)^3} 
		[& 2X_i^h  (1-\sum_{\substack{j=1 \\ j \neq h}}^3 q_j) (3q_h^2+3q_h(-1+\sum_{\substack{j=1 \\ j \neq h}}^3 q_j)  \nonumber \\
		&+(-1+\sum_{\substack{j=1 \\ j \neq h}}^3 q_j)^2)+2q_h^3(-1+ \sum_{\substack{j=1 \\ j \neq h}}^3 X_i^j)]
	\end{align*}
	
	Furthermore, its off-diagonal elements are determined by
		\[
	\psi_{\pmb{q},h}^{k,l}(\pmb{X}_i)=\frac{2(X_i^1+X_i^2+X_i^3-1)}{(1-q_1-q_2-q_3)^3},  \text{ for } k \ne h \text{ or } l \ne h.
	\]
		Note that, since $q_h \ne 0, \, h \in \{1,2,3\}$ and $1-q_1-q_2-q_3 \ne 0$, the function  $\pmb{q} \mapsto \pmb{\psi}_{\pmb{q}}(\pmb{x})$ is twice continuously differentiable.
	
	The condition $\mathrm{E}_{\pmb{q}_0} \pmb{\psi}_{\pmb{q}_0}(\pmb{X}_i)=0$ follows by direct calculation. Moreover, by using $q_{0,h} \leq \xi$ for $h \in \{1, \ldots, 3\}$, it can be shown that
	\[
	\mathrm{E}_{\pmb{q}_0}\lVert \pmb{\psi}_{\pmb{q}_0} (\pmb{X}_i) \rVert^2= \frac{1}{q_{0,1}}+\frac{1}{q_{0,2}}+\frac{1}{q_{0,3}}+\frac{3}{1-q_{0,1}-q_{0,2}-q_{0,3}} < \frac{6}{\xi} < \infty.
	\]

	Due to \eqref{deriHD} and \eqref{deriND}, $\mathrm{E}_{\pmb{q}_0} \dot{\pmb{\psi}}_{\pmb{q}_0}(\pmb{X}_i)
	= -(1-q_{0,1}-q_{0,2}-q_{0,3})^{-1}(\mathbf{I}_{3 \times 3} -diag (
		(q_{0,2}+q_{0,3}-1)/q_{0,1}, (q_{0,1}+q_{0,3}-1)/q_{0,2}, (q_{0,1}+q_{0,2}-1)/q_{0,3}))$, where $\mathbf{I}_{3 \times 3}$ contains only ones and $diag$ is a diagonal matrix. The matrix $\mathrm{E}_{\pmb{q}_0} \dot{\pmb{\psi}}_{\pmb{q}_0}(\pmb{X}_i)$ is nonsingular with determinant $- (q_{0,1}q_{0,2}q_{0,3}(1-q_{0,1}-q_{0,2}-q_{0,3}))^{-1} \neq 0$, and its inverse can be calculated, with the adjoint,s $- \pmb{\Sigma}$.
		The required $\ddot{\psi}(\pmb{x})$ as uniform bound for the log-likelihood is given in \ref{appb}. Thus, all requirements of \citet[][Theorem 5.41]{vaart1998} are fulfilled and the sequence $\sqrt{n}(\hat{\pmb{q}}_n- \pmb{q}_0)$ is asymptotically normal with mean zero and covariance matrix $\pmb{\Sigma}$. (This can also be seen by \citet[][Theorem 5.3.5(ii)]{bickel}.) Let $T$ be such a normal distributed random vector with mean zero and covariance matrix $\Sigma$, so that $\sqrt{n}(\hat{\pmb{q}}_n- \pmb{q}_0) \to T$ in distribution. Hence, by using the delta-method \cite[see][Chapt. 3]{vaart1998}, the sequence $\sqrt{n}(\hat{\pmb{\rho}}_n- \pmb{\rho}_0) = \sqrt{n}[\pmb{\phi}(\hat{\pmb{q}}_n)- \pmb{\phi}(\pmb{q}_0)]$ converges to $\pmb{\phi}_{\pmb{q}_0}' \cdot T$ in distribution. \qed

Again comparing with the na\"ive srs-design of intra-triplet independence, the variance calculates as $\mathrm{Var}(\hat{\pi}_n)_{srs}= \pi_0(1-\pi_0)/(3n)$ being estimable (see \eqref{pigleichpi}) as 
\begin{equation} \label{sesrs}
	\widehat{\mathrm{Var}}(\hat{\pi}_n)_{srs}=\frac{1}{3n}\hat{\pi}_n(1-\hat{\pi}_n). 
\end{equation}
An analytical comparison with the standard error in Corollary \ref{Cor1}, in order to assess the presumable design effect of a  'inflation' of the standard error, is difficult. However, we will see the effect now in the application. 

\section{Simulation and application to real data} \label{sec3}

In order to verify the consistency from Lemma \ref{lamm1} of estimating $\pi$, $\rho_1$ and $\rho_2$ by estimator $\pmb{\rho}_n$, numerically, we simulate mean squared errors (MSE) for different samples sizes. 
Therefore, as true parameter vector we use $\pmb{\rho}_{0}=(\pi_0, \rho_{1,0}, \rho_{2,0})^T=(0.1,0.1,0.1)^T$. 
Hence, in Bahadur representation \eqref{Bahadur} true values are $\pmb{q}_0=(q_{0,1}, q_{0,2}, q_{0,3})^T=\pmb{\phi}^{-1}(\pmb{\rho})=(0.2052, 0.0378, 0.0064)^T$, and $q_{0,0}=1-q_{0,1}-q_{0,2}-q_{0,3}=0.7506$. We simulate, $1000$ times, a multinomial random variable with $n=37$, to mimic our application later, and $100, 500, 1000$. Results are in Table  \ref{tabelle1} and demonstrate consistency for all three parameters, however at slower rate for the correlations, especially that of order two.  
\begin{table}\caption{Mean squared error for $\hat{\pmb{\rho}}_n$ of Lemma \ref{lamm1} for increasing sample size $n$}\label{tabelle1}
	\begin{center}
		\begin{tabular}[h]{l c c c r}
			\hline \hline
			$n$ & $37$ & $100$ & $500$ & $1000$ \\
			\hline
			$MSE_{\pi}$ & $0.0011$ & $0.0004$ & $6.96\cdot10^{-5}$ & $3.43\cdot10^{-5}$  \\
			
			$MSE_{\rho_1}$ & $0.0175$ & $0.0075$ & $0.0014$ & $0.0007$ \\
			
			$MSE_{\rho_2}$ & $0.0964$ & $0.0452$ & $0.0093$ & $0.0045$ \\
			\hline \hline
		\end{tabular}
	\end{center}
\end{table}

As an application to public dental health we estimate the proportion of children without caries from a sample of $n=37$ nursing schools \cite[for a data description see][]{weissbach2013}. To suit our model, we randomly select three children from each nursing school. Estimates $\hat{\pmb{\rho}}_n$, based on $\hat{\pmb{q}}_n$, are given in Table \ref{tabelle2}. For the standard errors, the unknown true $\pmb{q}_0$ in Corollary \ref{Cor1} is replaced by $\hat{\pmb{q}}_n$. 
\begin{table}\caption{Estimates of $\hat{\pmb{q}}_n$ and $\hat{\pmb{\rho}}_n$ for $n=37$ nursing schools, together with estimated standard errors by Corollary \ref{Cor1}. Comparison with srs-design for $\pi_0$ (see  \eqref{pigleichpi} and \eqref{sesrs}).}\label{tabelle2}
	\begin{center}
		\begin{tabular}[h]{ccc}
			\hline
			\hline
			$\hat{q}_{n,1}=0.081$ & $\hat{q}_{n,2}=0.297$ & $\hat{q}_{n,3}=0.595$ \\
			\hline\noalign{\smallskip}
			$\hat{\pi}_n^{\pm SE}=0.820^{\pm 0.042}$ & $\hat{\rho}_{1,n}^{\pm SE}=0.146^{\pm 0.133}$ & $\hat{\rho}_{2,n}^{\pm SE}=-0.168^{\pm 0.262}$ \\
			\hline\noalign{\smallskip}
			\multicolumn{3}{c}{$\hat{\pi}_n^{\pm SE_{srs}}=0.820^{\pm 0.036}$} \\
			\hline
			\hline
		\end{tabular} 
	\end{center}
\end{table}
First of all, the data is too small to identify intra-triplet dependence as significant. In contrast to the equal point estimates of $\pi_0$ in both designs (see \eqref{pigleichpi}), the standard error from the $37$ triplets is around 10\% larger than from $3 \cdot 37=111$ independent children (with variance \eqref{sesrs}).

\section{Discussion}

We study the saturated model for a simple sample of triplets. With one parameter less, the triplet count can assumed to be Negative Binomial to account for intra-triplet dependence by one over-dispersion parameter (see \citet[][Sect. 13.4]{Agresti} or \citet{weisradl2019}). 

Smaller models for triplets are one aspect, another practical aspect are equally sized clusters with more than three cluster units. In view of 
Corollary \ref{Cor1} formulae for standard errors will then be unmanageable. Even for a sample of triplets, matrices in Corollary \ref{Cor1}, and hence  standard errors, can be considered too tedious to compute. It appears important study further the conditions for a parametric bootstrap \cite[see][Lemma 23.3]{vaart1998}, which appear to be fulfilled. This is especially tempting because of the simple expression $\hat{\pi}_n=\phi_1(\hat{\pmb{q}}_n)$.
Further, for unequal clusters sizes, \cite{stefanescu2003} proposes to use the maximal cluster size and think of smaller clusters to miss according outcomes and replace with the EM-algorithm \cite[see e.g.][for similar applications]{kremweis2012,stroweis2015}. However, such maximal size $n$ is then random, which is not covered by the ML theory as presented in \cite{vaart1998}.

{\it Acknowledgement:} The financial support from the Deutsche Forschungsgemeinschaft (DFG) is
gratefully acknowledged (Grant 386913674 'Multi-state, multi-time,
multi-level analysis of health-related demographic events: Statistical
aspects and applications'). For generous support we thank M. Herzog, A. Dartsch and Brian Bloch.

\begin{appendix}

	\section{Uniform bound for Theorem \ref{theo1}} \label{appa}
We find function $g(\pmb{x})$ with
	$\lVert \psi_{\pmb{q}}(\pmb{x}) \rVert = \sqrt{\psi_{q,1}(\pmb{x})^2+\psi_{q,2}(\pmb{x})^2+\psi_{q,3}(\pmb{x})^2} \leq g(\pmb{x})$.
Considering the fact that all coordinates of the vector $\pmb{\psi}_{\pmb{q}}(\pmb{x})$ have the same structure, we can use the same upper bound for all components. By suppressing the subscript $i$, we arrive for $h \in \{1,2,3\}$ at
\begin{align}
	\psi_{\pmb{q},h}(\pmb{x})^2&=\biggl(\frac{1}{q_h(1-\sum_{j=1}^3q_j)}\Bigl((1-\sum_{\substack{j=1 \\ j \neq h}}^3 q_j)x_h+q_s(-1+\sum_{\substack{j=1 \\ j \neq h}}^3 x_j) \Bigl)\biggr)^2. \label{nachweis1}
\end{align}
Since $\pmb{q} \in \pmb{Q} \subseteq [\xi, 1-\xi]^3$, it is $1-\sum_{\substack{j=1 \\ j \neq h}}^3 q_j \leq 1-2\xi <1-\xi$ and $q_h\leq 1-\xi$. Furthermore, to minimize the denominator, we have $q_h \geq \xi$ and $1-\sum_{j=1}^3q_j \geq \xi$. Inserting the latter into \eqref{nachweis1}, it follows $\psi_{\pmb{q},h}(\pmb{x})^2<(1-\xi)^2(x_1+x_2+x_3-1)^2/\xi^4$. Consequently, $g(\pmb{x})$ can be defined as \eqref{hdef} due to
\begin{align*}
	\lVert \psi_{\pmb{q}}(\pmb{x}) \rVert < \sqrt{3 (1-\xi)^2(x_1+x_2+x_3-1)^2 \xi^{-4}}.
\end{align*}
Finally, in order to show integrability note that
\begin{align*}
	\mathrm{E}[g(\pmb{X})]=&\sqrt{3}(1-\xi) \xi^{-2} \mathrm{E}(X^1+X^2+X^3-1) \nonumber \\
	=&\sqrt{3}(1-\xi) \xi^{-2} (q_{0,1} \cdot 0+q_{0,2}\cdot 0+q_{0,3} \cdot 0-(1-q_{0,1}-q_{0,2}-q_{0,3})) \nonumber \\
	=&\sqrt{3}(1-\xi) \xi^{-2} (q_{0,1}+q_{0,2}+q_{0,3}-1) < \sqrt{3}(1-\xi) \xi^{-2} (3(1-\xi)-1) \nonumber \\
	=&\sqrt{3}(1-\xi) \xi^{-2} <\infty.
\end{align*}


\section{Uniform bound for Corollary \ref{Cor1}} \label{appb}
We here construct the bound for elements of the Hessian to the log-likelihood, called $\ddot{\psi}(\pmb{x})$, in \citet[][Theorem 5.41]{vaart1998}. Since all second order derivatives of the form $\psi_{\pmb{q},h}^{h,h}$, $h \in \{1,2,3\}$,  have the same structure and the off-diagonal elements of $\ddot{\pmb{\psi}}_{\pmb{q}}(\pmb{X}_i)$ are identical, we only have to construct bounds for these two cases. Due to the fact that $\pmb{q} \in \pmb{Q} \subseteq [\xi, 1-\xi]^3$, the function $g_1(\pmb{x})$ with $|\psi_{\pmb{q},1}^{1,1}(\pmb{x})|<g_1(\pmb{x})$ can be defined by:
\begin{eqnarray*}
	|\psi_{\pmb{q},1}^{1,1}(\pmb{x})| &=& |2x_1(1-q_2-q_3)(3q_1^2+3q_1(q_2+q_3-1)+(q_2+q_3-1)^2) \\
	& & +2q_1^3(x_2+x_3-1)|/|q_1^3(1-q_1-q_2-q_3)^3| \\
	&< & \frac{1}{\xi^6}( 2|x_1|(1-2\xi)(3(1-\xi)^2+3(1-\xi)(1-2\xi)+(1-2\xi)^2)+ \\
	& & 2(1-\xi)^3|(x_2+x_3-1)|)  \\
	&<& |\psi_{\pmb{q},1}^{1,1}(\pmb{x})|<\frac{1}{\xi^6}\bigl( 2(1-\xi)^3(7|x_1|+|x_2+x_3-1|)\bigr)=:g_1(\pmb{x})
\end{eqnarray*}
Analogously, we get $g_2(\pmb{x}):= \frac{1}{\xi^6}[2(1-\xi)^3(7|x_2|+|x_1+x_3-1|)]$ and  $g_3(\pmb{x}):= \frac{1}{\xi^6}[2(1-\xi)^3(7|x_3|+|x_1+x_2-1|)]$. For the off-diagonal elements ($l_1 \neq h \lor l_2 \neq h$), it is
\begin{align*}
	|\psi_{\pmb{q},h}^{l_1,l_2}(\pmb{x})| = \Bigl| \frac{2(x_1+x_2+x_3-1)}{(1-q_1-q_2-q_3)^3}\Bigr| \leq \frac{2}{\xi^3}|x_1+x_2+x_3-1| =: g(\pmb{x}).
\end{align*}
Finally we can define $\ddot{\psi}(\pmb{x}):= \max(g_1(\pmb{x}), g_2(\pmb{x}), g_3(\pmb{x}), g(\pmb{x}))$. Note that, due to its continuity, $\ddot{\psi}(\pmb{x})$ is integrable.

\end{appendix}

\end{document}